\documentclass[12pt,amsmath,prl,showpacs,superscriptaddress,subeqn]{revtex4}

\usepackage{amsmath}
\usepackage{amsfonts}
\usepackage{graphicx}
\usepackage{color}

\begin{document}

\title{Stimulated Raman Adiabatic Passage (STIRAP) as a Route to Achieving Optical Control in Plasmonics}
\date{\today}
\author{Maxim Sukharev}
\email{maxim.sukharev@asu.edu}
\affiliation{Department of Applied Sciences and Mathematics, Arizona State University, Mesa, Arizona 85212, USA}
\author{Svetlana A. Malinovskaya}
\affiliation{Department of Physics and Engineering Physics, Stevens Institute of Technology, Hoboken, New Jersey 07030, USA}

\begin{abstract}
Optical properties of ensembles of three-level quantum emitters coupled to plasmonic systems are investigated employing a self-consistent model. It is shown that stimulated Raman adiabatic passage (STIRAP) technique can be successfully adopted to control optical properties of hybrid materials with collective effects present and playing an important role in light-matter interactions. We consider a core-shell nanowire comprised of a silver core and a shell of coupled quantum emitters and utilize STIRAP scheme to control scattering efficiency of such a system in a frequency and spatial dependent manner. After the STIRAP induced population transfer to the final state takes place, the core-shell nanowire exhibits two sets of Rabi splittings with Fano lineshapes indicating strong interactions between two different atomic transitions driven by plasmon near-fields.
\end{abstract}


\maketitle

\section{Introduction}\label{intro}
The research field of plasmonics, while still expanding its applications in linear nano-optics \cite{MaierAtwater05,BarnesMurray07,Berini09,Moreno09,GramotnevBozhevolnyi10,StockmanOptExpr}, is quickly advancing towards nonlinear phenomena \cite{PanoiuOsgood04,SchatzGray06,Czitrovszky08}. Recently, it has been proposed to combine plasmonic systems with highly nonlinear media \cite{Zheludev06,Zayats06,Psaltis10}. Utilizing strongly inhomogeneous electromagnetic (EM) fields associated with the surface plasmon-polariton (SPP) resonance, one can achieve a significant spatial dependence of density of the conductive electrons in metals resulting in nonlinear phenomena such as second harmonic generation \cite{Stockman04,Moloney09}. Extreme concentration of EM radiation was proposed to be utilized as a catalyst to achieve lasing in nano-systems and has been recently implemented in experiments \cite{Zhang09}. Other important applications of the light localization include surface enhanced Raman spectroscopy (SERS) \cite{EtchegoinSERS} and solar energy harvesting \cite{Polman08,Atwater10}.

A quickly growing field of hybrid materials is emerging \cite{Artemyev07,SalomonEbbesen09,Richards09,Gray11,Park12} on the base of latest advancements in nanoplasmonic science. Here one merges plasmonics with atomic and molecular physics considering systems comprised of quantum emitters and metal nano-structures. With current advances in chemistry and nano-manufacturing one is now able to couple molecular ensembles to plasmonic systems. Such systems exhibit a wide variety of new phenomena including new mixed molecular-plasmon states \cite{SalomonSukharev12} and plasmon control of molecular energy redistribution  \cite{Bouhelier07}.

It has long been realized that ideas of coherent control developed in quantum chemistry and physics \cite{Shapiro} could be successfully applied to optically active nano-systems \cite{Stockman02,Gray05,Sukharev06,Brixner07,Giessen10} controlling electron transport \cite{Fainberg11}, light pathways \cite{Sukharev07}, and EM hot spots \cite{Zheludev11}. The ultimate goal in these investigations is to achieve control of optical properties of nano-structures.

This paper explores ideas of stimulated Raman adiabatic passage (STIRAP) applied to ensembles of three-level atoms optically coupled to plasmonic systems. STIRAP is known to be based on the adiabatic population transfer within a single dressed state that does not include the dark transitional state thus minimizing spontaneous losses. The scheme has a variety of attractive modern applications from cooling internal degrees of freedom in molecules \cite{Ye08}, to maximizing coherence between the initial and final states \cite{ScullyFracSTIRAP}, to manipulating dynamics in a multilevel system by making use of the Optimal Control Theory that reveals STIRAP type of control \cite{KumarOCTSTIRAP}. The goal of the paper is twofold - first, to analyze the efficiency of STIRAP technique in the ensemble of emitters where collective effects are taken into account in the framework of Maxwell-Liouville-von Neumann equations, and, second, to demonstrate the implementation of STIRAP as a tool to control scattering, reflection, and transmission properties of hybrid systems. As an example we consider a core-shell silver nanowire with resonantly coupled layer of three-level atoms.

\section{Model}\label{model}
We consider electrodynamics of ensembles of quantum emitters in a self-consistent approach. We solve the system of Maxwell's equations in time domain for electric, $\vec{E}$, and magnetic, $\vec{H}$, fields. In spatial regions occupied by quantum emitters the Maxwell equations read
\begin{align}
\label{Maxwell}
\varepsilon_0\frac{\partial\vec{E}}{\partial t}=\nabla\times\vec{H}-\frac{\partial\vec{P}}{\partial t},
\\
\mu_0\frac{\partial\vec{H}}{\partial t}=-\nabla\times\vec{E},
\nonumber
\end{align}
where $\varepsilon_0$ and $\mu_0$ and are the dielectric permittivity and the magnetic permeability of the free space, respectively, and $\vec{P}$ is the macroscopic polarization of a quantum medium. The latter is calculated using the mean-field approximation
\begin{equation}
\label{polarization}
\vec{P}=n_a \langle \vec{d}\rangle,
\end{equation}
where $\langle \vec{d}\rangle=$Tr$(\hat{\rho}\vec{d})$ is the expectation value of quantum emitter's dipole moment. The dynamics of the density matrix $\rho$ satisfies the Liouville-von Neumann equation
\begin{equation}
\label{Liouville}
i\hbar\frac{d\hat{\rho}}{dt}=[\hat{H},\hat{\rho}]-i\hbar\hat{\Gamma}\left (\hat{\rho} \right)\hat{\rho},
\end{equation}
here $\hat{H}$ is the Hamiltonian that describes the atom-EM field interaction. We assume that the relaxation processes are Markovian. The superoperator $\hat{\Gamma}$ accounts for the decay of the excited state and dephasing effects.

In the mean-field approximation employed here it is assumed that the density matrix of the atomic ensemble is expressed as a product of density matrices of individual atoms (\ref{Liouville}) driven by a local EM field (\ref{Maxwell}). In order to take into account dipole-dipole interactions of atoms within a single grid cell we follow Ref. \cite{DowlingPRA93} and introduce Lorentz-Lorenz correction term for a local electric field according to
\begin{equation}
\label{LL}
\vec{E}_{\text{local}}=\vec{E}+\frac{\vec{P}}{3\varepsilon_0},
\end{equation}
where $\vec{E}$ is the solution of Maxwell's equations (\ref{Maxwell}) and macroscopic polarization is evaluated according to Eq. (\ref{polarization}). We performed several test simulations comparing results with and without local field correction term Eq. (\ref{LL}). It was found that frequency dependencies of observables, for example,  the transmission coefficient is affected by Eq. (\ref{LL}) at high atomic densities resulting in slight changes of resonant frequencies. If one is interested in qualitative analysis, such variations are not important. However, for the sake of completeness we use Eq. (\ref{LL}) in all simulations below.

The system of equations (\ref{Maxwell}) and (\ref{Liouville}) coupled via (\ref{polarization}) with local field correction (\ref{LL}) is solved on a parallel multiprocess cluster following the numerical algorithm discussed in Ref. \cite{SukharevNitzan11}.

\section{Results and discussion}\label{results}
\begin{figure}[tbph]
\centering\includegraphics[width=\linewidth]{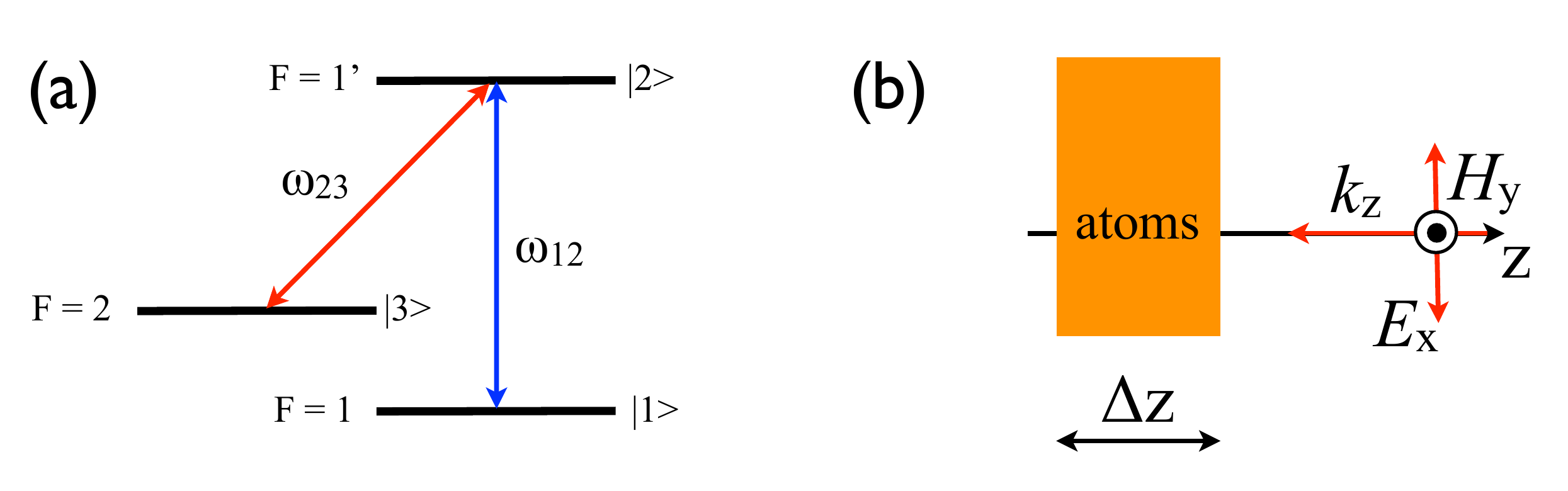}
\caption{(Color online) One-dimensional STIRAP. Panel (a): the energy diagram of a three-level atom with two dipole transitions indicated as blue and red arrows. Panel (b): a schematic setup of simulations with a layer of atoms of the thickness of $\Delta z$ exposed to the incident field propagating along negative $z$ direction and polarized vertically. In all one-dimensional simulations the following set of parameters is used: $\omega_{12}=2.9$ eV, $\omega_{23}=2.8$ eV, $\Delta z=200$ nm.}
\label{fig1}
\end{figure}

First, we consider a $\Lambda$-system with the energy level diagram depicted in Fig. \ref{fig1}a describing an atom in the ensemble of interacting emitters. An infinite in $x$ and $y$ dimensions and finite in $z$ dimension quantum medium with the thickness of $\Delta z$ is driven by an incident linearly-polarized field. The vector of light polarization is along $x$ axis and the light propagates along negative $z$ axis as shown in Fig. \ref{fig1}b. 

The Hamiltonian of a single three-level atom reads
\begin{equation}
\label{1D-Hamiltonian}
\hat{H}=\hat{H_0}-\vec{\mu}\vec{E}_{\text{local}}=
\left(
\begin{array}{ccc}
0 & 0 & -\frac{E_{{\text{local}},x}\mu_{12}}{\sqrt6} \\
0 & \hbar\omega_{23} & -\frac{E_{{\text{local}},x}\mu_{23}}{\sqrt10} \\
-\frac{E_{{\text{local}},x}\mu_{12}}{\sqrt6} & -\frac{E_{{\text{local}},x}\mu_{23}}{\sqrt10} & \hbar\omega_{12}
\end{array}
\right),
\end{equation}
where $\mu_{12}$ and $\mu_{23}$ are transition dipoles (see Fig. \ref{fig1}a). In all simulations we use $\mu_{12}=\mu_{23}=2$ Debye. The Hamiltonian is written in the basis of states of angular momentum ($\left|1\right>$, $\left|3\right>$, $\left|2\right>$), the coefficients $1/\sqrt{6}$ and $1/\sqrt{10}$ are the Clebsch-Gordan coefficients.

\begin{figure}[tbph]
\centering\includegraphics[width=\linewidth]{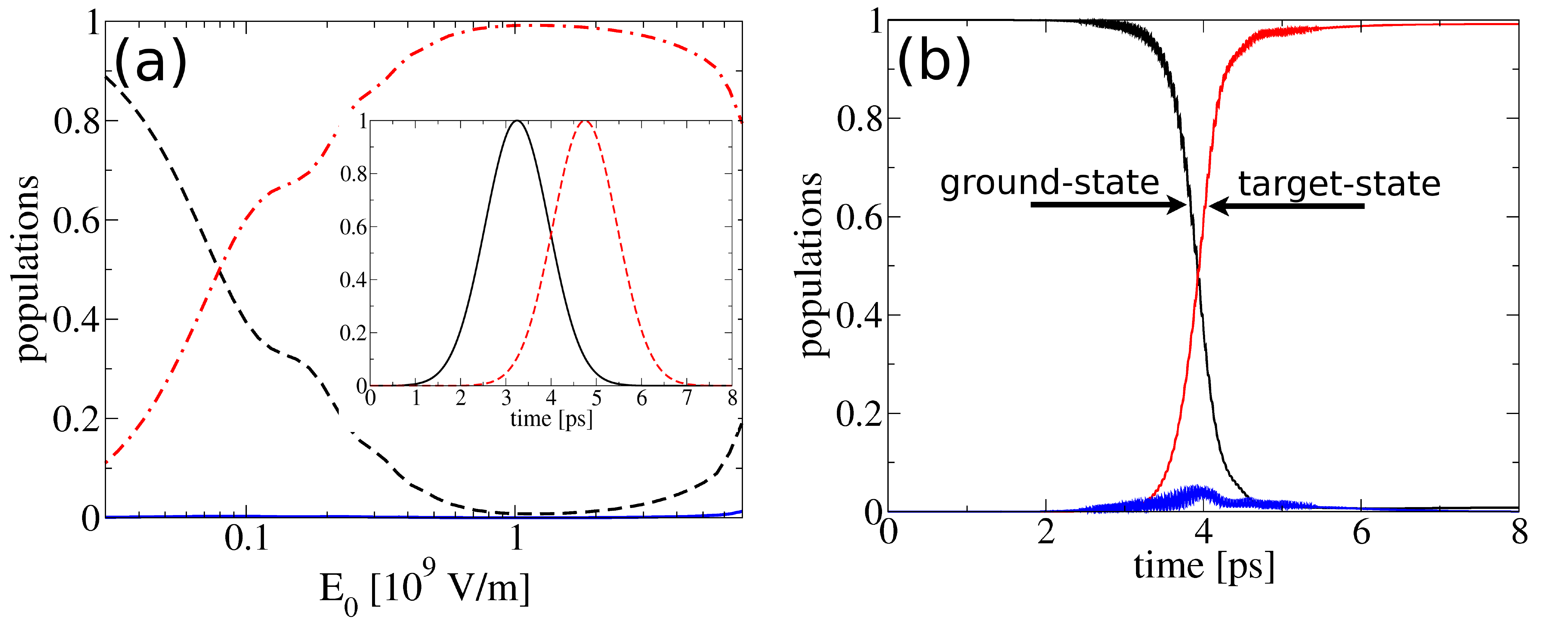}
\caption{(Color online) Single atom STIRAP. Panel (a): atomic state populations as functions of the incident field amplitude $E_0$ (in V/m) for a single atom case. Ground state, $\left| 1 \right>$, population is shown as a dashed black line, target state, $\left| 3 \right>$, population is indicated as a dash-dotted red line, the solid blue line shows the intermediate state, $\left| 2 \right>$, population. The inset depicts STIRAP scheme showing Stokes (solid black line) and pump (dashed red line) pulses as functions of time in ps. Panel (b): time dynamics of the atomic populations during STIRAP pulse sequence (here the color scheme is the same as in panel (a)). The parameters for STIRAP pulses (see eq. (\ref{STIRAP})) are: $E_{01}=E_{02}=E_0=1.1\times10^9$ V/m, $\tau_1=\tau_2=1$ ps, $\Delta \tau=1.5$ ps, $t_0=4$ ps.}
\label{fig2}
\end{figure}

As initial conditions we use an incident field in the form
\begin{align}
\label{STIRAP}
E_{x,\text{inc},1}=E_{01}\cos(\omega_{12}( t-\frac{\Delta\tau}{2}-t_0))\exp(-\frac{(t-\frac{\Delta\tau}{2}-t_0)^2}{\tau_1^2}),
\\
E_{x,\text{inc},2}=E_{02}\cos(\omega_{23}( t+\frac{\Delta\tau}{2}-t_0))\exp(-\frac{(t+\frac{\Delta\tau}{2}-t_0)^2}{\tau_2^2}),
\nonumber
\end{align}
where indexes $1$ and $2$ correspond to the pump field (to pump an atom from its ground state $\left|1\right>$ to the excited state $\left|2\right>$) and the Stokes field (to create a superposition of the $\left|2\right>$ and the $\left|3\right>$ states), with the latter preceding the former in time by $\Delta\tau$. Other parameters are the pulse duration $\tau$ and the central time $t_0$. Their values are presented in the caption of Fig. \ref{fig2}.

Our initial goal is to examine how STIRAP scheme for  a single atom is affected by mutual EM interactions of atoms. In order to have reference data we first perform a series of simple calculations for a single atom case summarized in Fig. \ref{fig2}. The excitation pulse sequence is shown in the inset of Fig. \ref{fig2}a. To determine optimal amplitudes of the Stokes and pump pulses, $E_{01}$, $E_{02}$ we calculate populations of atomic levels at the end of the pulse sequence scanning through their peak values $E_{01}$ and $E_{02}$ (here for simplicity we assume that $E_{01} = E_{02}$). We chose the pulse duration $\tau_1$ and $\tau_2$ to be such that $1/ \tau_i \le (\omega_{12} - \omega_{23})= \omega_{13}$ to resolve spectrally the splitting between the initial and final states. Fig. \ref{fig2}a shows that the optimal STIRAP occurs at $E_{01}=E_{02}=1.1\times10^9$ V/m with the population of the target state $\left| 3 \right>$ reaching $0.992$ while the ground state population is $8\times10^{-3}$ and the state $\left| 2 \right>$ is nearly $0$. In these calculations, the pulse duration $\tau$ is chosen to be 1 ps, giving $1/\tau = 5 \omega_{13}$. Note that longer pulse duration provides higher efficiency of population transfer up to $100\%$, our choice of the value of the $\tau$ is dictated by computational reasons. Fig. \ref{fig2}b shows time dynamics in a single atom under the optimal STIRAP condition confirming that the chosen scheme follows the conventional STIRAP.

\begin{figure}[tbph]
\centering\includegraphics[width=\linewidth]{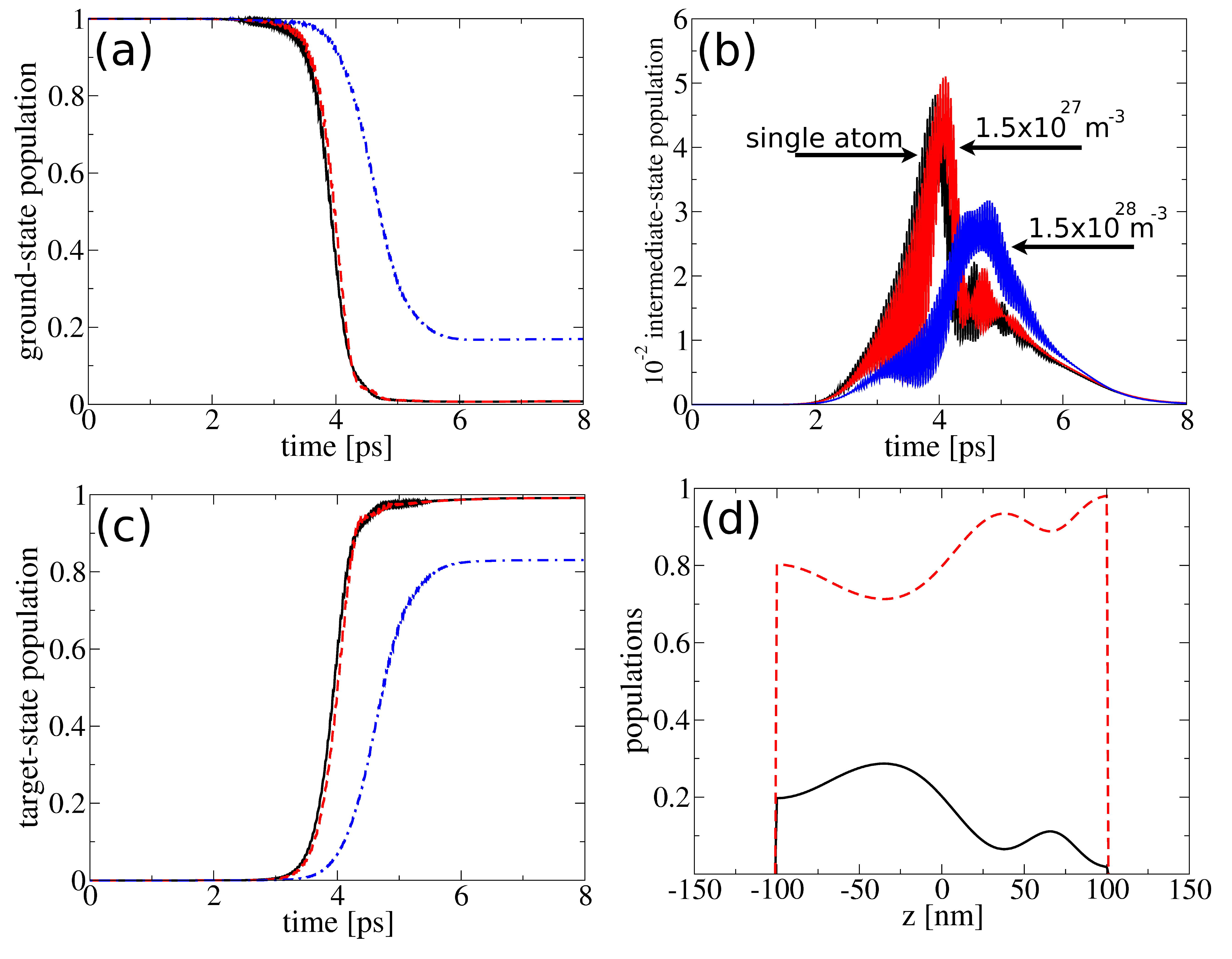}
\caption{(Color online) One-dimensional self-consistent calculations. Panel (a): spatially averaged ground state population as a function of time during STIRAP for a single atom case (solid black line), atomic layer at $n_a=1.5\times10^{27}$ m$^{-3}$ (dashed red line), and atomic layer at $n_a=1.5\times10^{28}$ m$^{-3}$ (dash-dotted blue line). Panel (b): same as in panel (a) but for the intermediate state. Panel (c): same as in panels (a) and (b) but for the target state. Panel (d): spatial distribution of the ground state (solid black line) and target state (dashed red line) populations after STIRAP as functions of the coordinate $z$ in nm at $n_a=1.5\times10^{28}$ m$^{-3}$. The pure dephasing rate is $0$, the radiationless decay rate for both atomic transitions is $10^{12}$ s$^{-1}$. The transition frequencies for the atomic system are $\omega_{21}=2.8$ eV, $\omega_{32}=2.9$ eV.}
\label{fig3}
\end{figure}

Next, we compare the results for a single atom with that obtained using one-dimensional self-consistent model for a layer of three-level atoms with thickness of $\Delta z=200$ nm. Panels (a) through (c) of Fig. \ref{fig3} show the time dynamics of atomic populations averaged over the layer's volume at different atomic densities, $n_a$, under the optimal STIRAP condition described above. At the density of $n_a=1.5\times10^{27}$ m$^{-3}$ the dynamics is nearly identical to that of a single atom. However at higher density of $n_a=1.5\times10^{28}$ m$^{-3}$ one can clearly see a noticeable difference. The final population of the target state degrades from its optimal value of $0.992$ to $0.831$ (Fig. \ref{fig3}c). It is interesting to note that even though STIRAP dynamics is significantly affected by strong coupling between atoms such that the intermediate state is populated during transitional times (Fig. \ref{fig3}b), it is still negligibly small at the end of the excitation. One can examine a spatial distribution of both the ground and the target states at the end of STIRAP pulse sequence as shown in Fig. \ref{fig3}d. Note that the incident field propagates from right to left (see Fig. \ref{fig1}b for details). Atomic state populations exhibit spatial modulations with higher population transfer at the input side of the layer. This suggests that the pump and Stokes field carrier frequency gets modulated as the fields propagate through the medium. The degree of modulation is different for the Stokes and pump frequencies which results in a deviation from the two-photon resonance condition and, thus, in reduction of the efficiency of STIRAP population transfer. Another maximum of the target state population is seen near $z=37$ nm with the corresponding minimum of the ground state population. STIRAP conditions, under which the sample is illuminated, are clearly affected by atom-atom interactions that alter incident field via spatially dependent stimulated EM radiation.

\begin{figure}[tbph]
\centering\includegraphics[width=\linewidth]{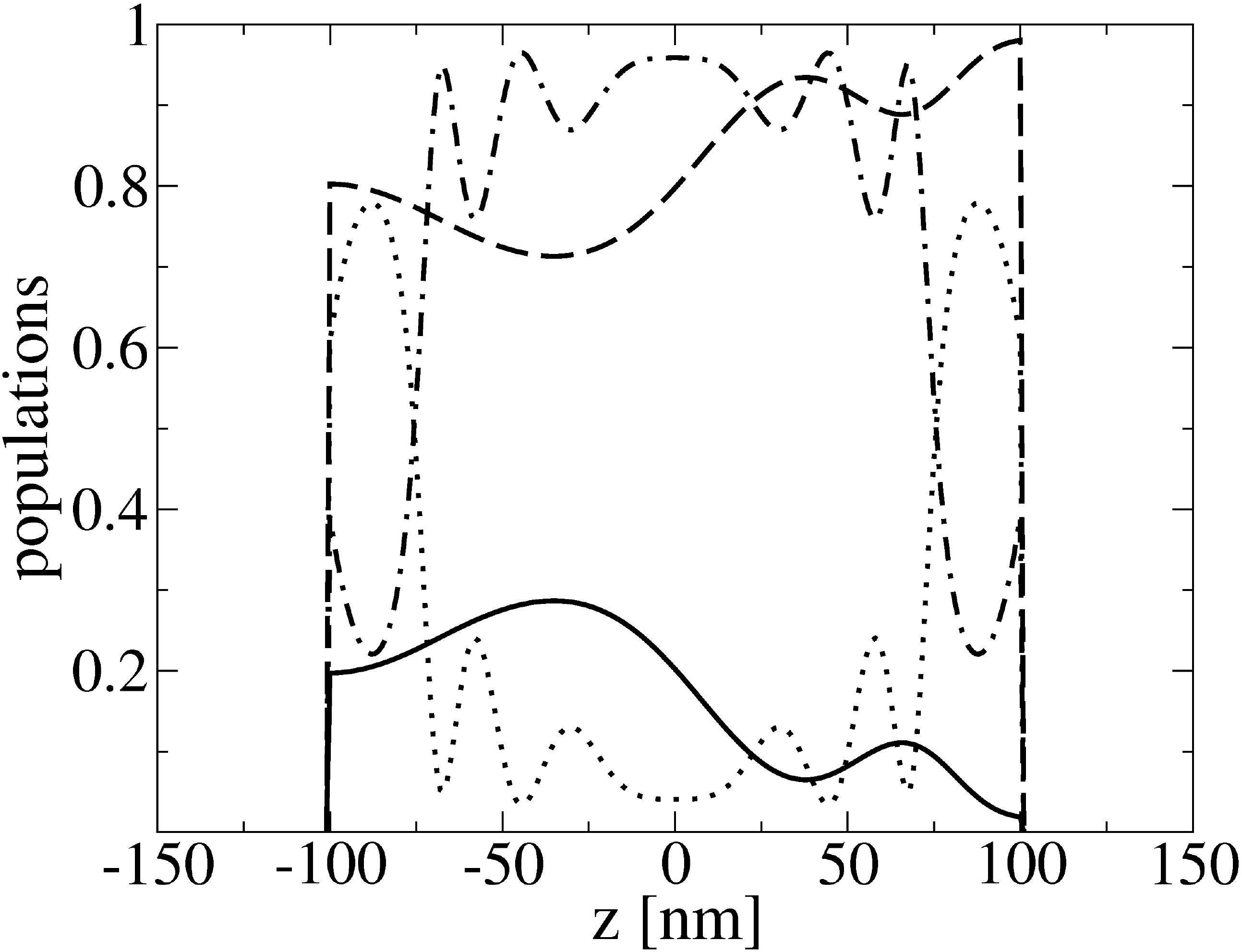}
\caption{One-dimensional self-consistent calculations. Spatial distribution of the ground state and target state populations after STIRAP as functions of the coordinate $z$ in nm at $n_a=1.5\times10^{28}$ m$^{-3}$ for two STIRAP excitation schemes: ground state population after a single-ended excitation - solid line, target state population after a single-ended excitation - dashed line, ground state population after the double-ended symmetric STIRAP - dotted line, target state population after the double-ended symmetric STIRAP - dash-dotted line. Other parameters are the same is in Fig. \ref{fig3}.}
\label{fig4}
\end{figure}

One may send two sets of STIRAP pulse sequences from both sides of the sample and examine the spatial dependence of atomic populations as shown in Fig. \ref{fig4}. Here we compare a single-sided excitation with a double-sided one. Several new features are noticed: a). the target state population becomes significantly higher in the center of the sample under symmetric double-sided STIRAP condition; b). spatial modulations of atomic state populations are more pronounced.

\begin{figure}[tbph]
\centering\includegraphics[width=\linewidth]{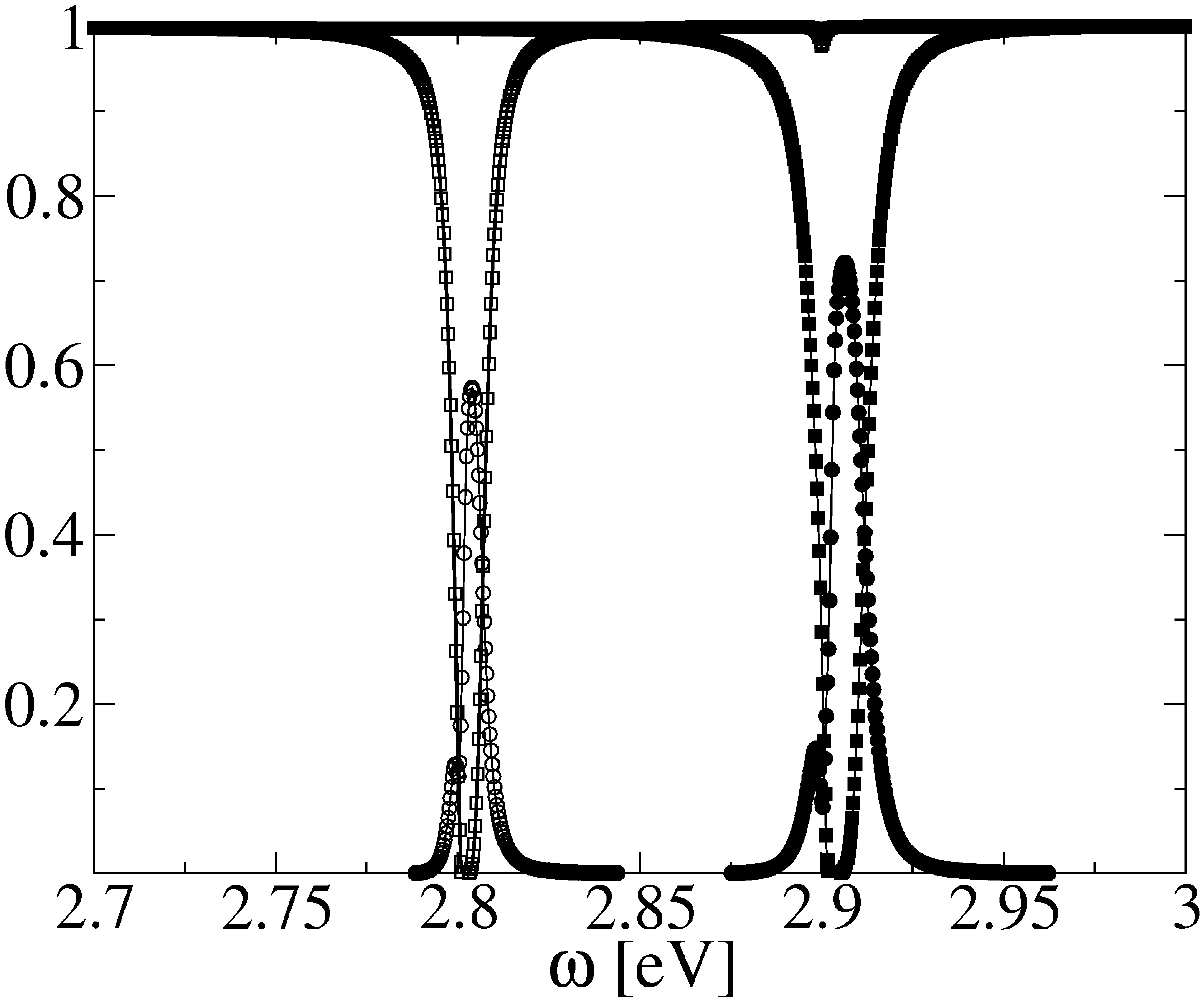}
\caption{One-dimensional STIRAP control. Reflection (circles) and transmission (squares) coefficients as functions of incident frequency, $\omega$, in eV for the atomic layer at $n_a=1.5\times10^{27}$ m$^{-3}$ before STIRAP (filled circles and squares) and after STIRAP (empty circles and squares).}
\label{fig5}
\end{figure}

To illustrate one of many applications of the STIRAP scheme for optical control we perform a series of simulations calculating transmission, $T$, and reflection, $R$, coefficients of an atomic layer before and after STIRAP. Fig. \ref{fig5} shows both sets of data. Before STIRAP pulse sequence partially inverts the atoms, the system has two reflection maxima near the transition frequency $\omega_{12}$ with $T$ exhibiting a wide minimum. The appearance of the second resonance in the reflection spectrum is a clear indication of a strong collective interaction of atoms in the layer that leads to appearance of new EM modes with frequencies other than $\omega_{12}$. A number of these modes increases with the increase of atomic density \cite{Glauber2010,Glauber2011}. After the STIRAP population transfer to the final state, which is not coupled to the ground state, both $R$ and $T$ have extrema near $\omega_{23}$ transition frequency. It should be noted that after STIRAP is applied, the transmission coefficient has an additional, small minimum at $\omega_{12}$. This is due to the fact that not all atoms are inverted and their small fraction still produces that resonance.

\begin{figure}[tbph]
\centering\includegraphics[width=\linewidth]{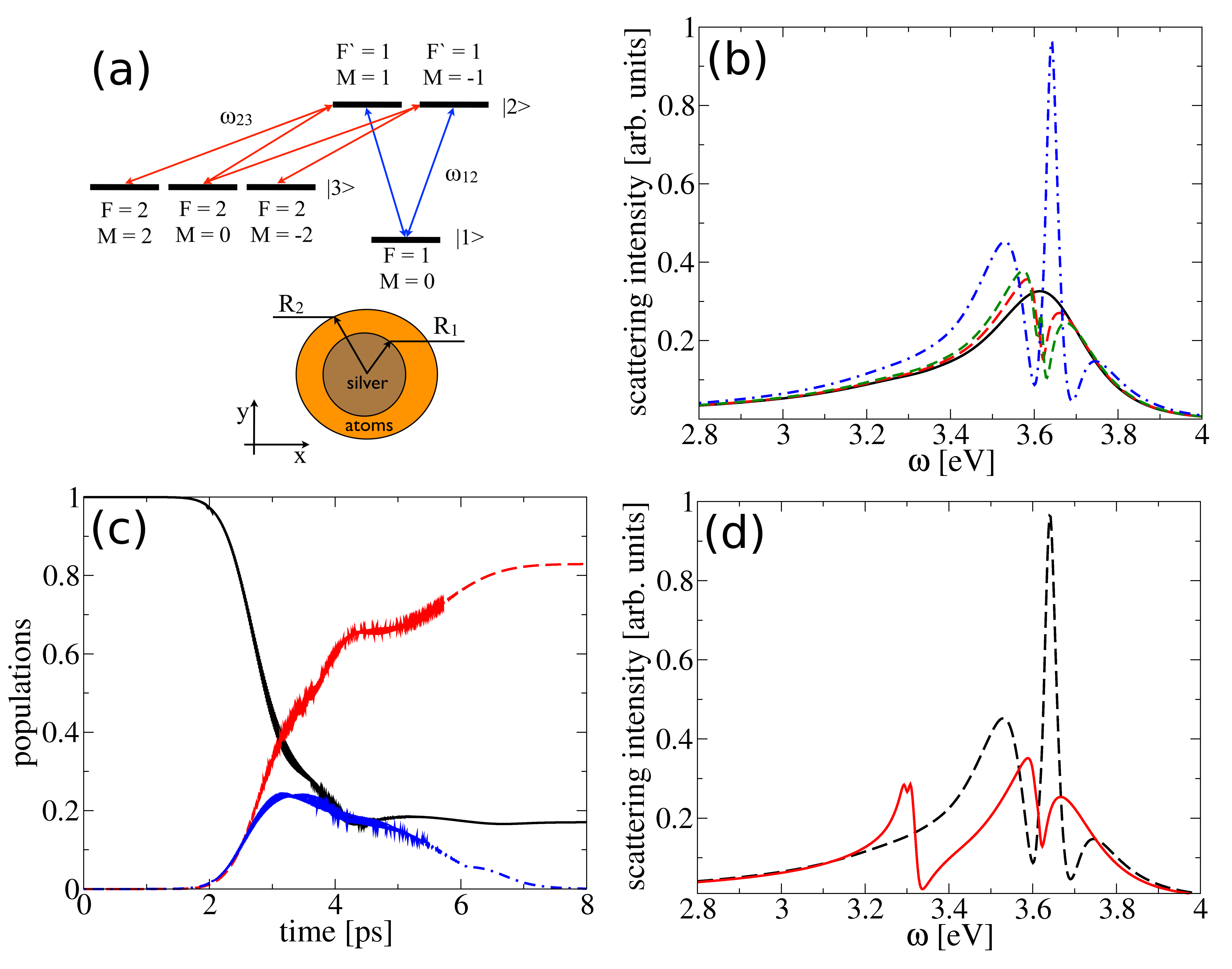}
\caption{(Color online) Two-dimensional STIRAP control of core-shell nanowires. Panel (a): energy level diagram of a three-level atom in two dimensions. In all two-dimensional simulations $\omega_{12}=3.61$ eV, $\omega_{23}=3.3$ eV. The schematic setup of a core-shell nanowire is shown below the energy diagram. In simulations radii of the core and shell are set at $R_1=20$ nm and $R_2=35$ nm, respectively. Panel (b): scattering intensity as a function of incident frequency, $\omega$, in eV. The scattering spectrum of a bare silver wire is shown as a solid black line. Long-dashed red line shows data for core-shell wire at the atomic density of $n_a=5\times10^{26}$ m$^{-3}$, short-dashed green line is for $n_a=10^{27}$ m$^{-3}$, and dash-dotted blue line is for $n_a=5\times10^{27}$ m$^{-3}$.  Panel (c): time dynamics of local atomic populations near surface of the sliver core during STIRAP. Ground state population - solid black line, intermediate state population - dash-dotted blue line, the target state population - dashed red line. Amplitudes of both Stokes and pump pulses is $3.4\times10^9$ V/m. The pure dephasing rate is $10^{13}$ s$^{-1}$, the radiationless decay rate for both atomic transitions is $10^{12}$ s$^{-1}$. Panel (d): same as in panel (b) but before STIRAP (dashed black line) and after STIRAP (solid red line) at $n_a=5\times10^{27}$ m$^{-3}$.}
\label{fig6}
\end{figure}

The second set of simulations is performed for a core-shell silver nanowire covered by a thin layer of three-level atoms as shown in Fig. \ref{fig6}a. Here we perform two-dimensional simulations, assuming that the silver nanowire extends infinitely in $z$ dimension. The system is excited by an incident electric field polarized along $x$ axis that propagates along $y$ axis.

The energy level diagram of the system is created on an example of the alkali atom, e.g., Rb. Here, the electron transitions are considered that are induced between the hyperfine states of the $5^{2}S_{1/2}$ and $5^{2}P_{1/2}$ electronic  states within D1 line of  $^{87}Rb$ (nuclear spin I=3/2). For the $5^{2}S_{1/2}$ state, $F$ can take values $2$ or $1$, and for the D1 excited state $5^{2}P_{1/2}$, $F$ is either $1$ or $2$. In our scheme, we choose the initial state to be $F=1$ of the $5^{2}S_{1/2}$ state and  the final state to be $F=2$ of the $5^{2}S_{1/2}$ state. The lowest hyperfine sate $F=1$ of the $5^{2}P_{1/2}$ is the transitional excited state \cite{Malinovskaya12}. We consider the population initially to be in state $F=1$ with the projection $M=0$ which we may prepare by optical pumping in the presence of the constant magnetic field that removes the degeneracy of the magnetic sub-levels. Typical transition frequencies in plasmonic materials are on the order of $1-4$ eV. In order to efficiently couple our model atom to SPP resonances we assume atomic transition frequencies to be on the same order as SPP modes.

The Hamiltonian of a single three-level atom in the two-dimensional nanowire geometry shown in Fig. \ref{fig6}a is
\begin{equation}
\label{2D-Hamiltonian}
\hat{H}=
\left(
\begin{array}{cccccc}
0 & 0 & 0 & 0 & -\frac{\Omega_+\mu_{12}}{2\sqrt3} & -\frac{\Omega_-\mu_{12}}{2\sqrt3} \\
0 & \hbar\omega_{23} & 0 & 0 & -\frac{\Omega_-\mu_{23}}{\sqrt10} & 0 \\
0 & 0 & \hbar\omega_{23} & 0 & -\frac{\Omega_+\mu_{23}}{2\sqrt15} & \frac{\Omega_-\mu_{23}}{2\sqrt15} \\
0 & 0 & 0 & \hbar\omega_{23} & 0 & -\frac{\Omega_+\mu_{23}}{\sqrt10} \\
-\frac{\Omega_-\mu_{12}}{2\sqrt3} & -\frac{\Omega_+\mu_{23}}{\sqrt10} & -\frac{\Omega_-\mu_{23}}{2\sqrt15} & 0 & \hbar\omega_{12} & 0 \\
-\frac{\Omega_+\mu_{12}}{2\sqrt3} & 0 & \frac{\Omega_+\mu_{23}}{2\sqrt15} & -\frac{\Omega_-\mu_{23}}{\sqrt10} & 0 & \hbar\omega_{12}
\end{array}
\right),
\end{equation}
where $\Omega_{\pm}=E_{{\text{local}},x}\pm i E_{{\text{local}},y}$.

It is informative first to examine scattering efficiency of such a system in the linear regime, when $\left| 1 \right>$ to $\left| 2 \right>$ absorption line is dominant and is in resonance with a SPP mode of the silver nanowire. Under such conditions the localized SPP resonance in the scattering spectrum splits into two modes, upper and lower polaritons, the phenomenon known as Rabi splitting. The scattering spectrum for the bare silver nanowire and core-shell system are shown in  Fig. \ref{fig6}b as functions of the incident frequency. The Rabi splitting is observed to increase with the density of the atomic ensemble. It is important to note that at high density of $n_a = 5\times10^{27}$ m$^{-3}$ we observe additional resonance - the collective atom-plasmon mode. The latter has been detected several times in recent experiments \cite{SalomonEbbesen09}. The physics of this mode was recently scrutinized in Ref. \cite{SalomonSukharev12}, where it was demonstrated that this mode is due to plasmon induced dipole-dipole interactions between the quantum emitters.

We apply the STIRAP excitation scheme with the parameters shown in the caption of Fig. \ref{fig6} to the atomic layer covering silver nanowire as in Fig. \ref{fig6}a. Time dynamics of atomic populations is shown in Fig. \ref{fig6}c. We note that nearly perfect STIRAP, observed in a single atom case, is suppressed in the atomic layer. This is owing to several factors. First is a fast decoherence whose rate is chosen to be $10^{13}$ s$^{-1}$, which is one order of magnitude smaller than the peak Rabi frequency. (Here, we considered two main channels of decoherence, spontaneous emission and collisional dephasing.) The second factor reducing the efficiency of STIRAP relates to the spacial features of the sample: different locations of the atomic layer are exposed to different EM fields due to spatially dependent strong local field enhancement near the surface of the silver core. Hence atoms in the shell are excited in a spatially dependent manner diminishing STIRAP. We performed additional simulations varying decoherence rates (within femtosecond time scale) and the atomic density (on the order of $10^{26}$ to $10^{28}$ m$^{-3}$). In all simulations STIRAP was suppressed but still quite noticeable as in Fig. \ref{fig6}c.

To utilize STIRAP technique as a possible control technique of optical properties of an atomic ensemble coupled to a plasmonic material, we calculate scattering intensity of the core-shell nanowire after STIRAP is complete. The results are shown in Fig. \ref{fig6}d for the density of $n_a = 5\times10^{27}$ m$^{-3}$. Even though obviously not all atoms are inverted to the target state, the spectrum appreciably differs from the one corresponding to all atoms in the ground state. We note several important features: a). the Rabi splitting, that is $216$ meV before STIRAP, is reduced to $78$ meV; b). the collective atom-plasmon mode is no longer seen; c). a new resonance near transitional atomic frequency $\omega_{23}$ is observed due to the presence of inverted atoms. An  additional resonant feature characteristic to inverted atoms is observed in the form of a tiny but distinct Rabi splitting in $3$ meV. This is due to the fact that the SPP resonance of the silver nanowire is very broad and can be coupled to both $\left| 1 \right>$ to $\left| 2 \right>$ and $\left| 2 \right>$ to $\left| 3 \right>$ atomic transitions. Moreover the strong interaction between these transitions is also seen in the spectrum. Both resonances have evident Fano lineshapes, which indicates coherent interactions between the atoms that are in different states.

\section{Conclusions}\label{conclusion}
In the framework of Liouville von Neuman equation coupled to the Maxwell equations within the self-consistent approach, we demonstrated that STIRAP technique may be used to control optical properties of ensembles of quantum emitters coupled to plasmonic materials. The importance of such control parameter as the atomic density is emphasized. The results are obtained using self-consistent calculations where the STIRAP scheme is explored in one and two dimensions taking into account collective effects. It is shown that at low densities STIRAP scheme gives the result nearly identical to that for a single atom. Simulations at higher densities revealed the significance of collective interactions between atoms that eventually diminish the STIRAP control mechanism. When STIRAP scheme was applied  to hybrid nano-structures comprised of coupled three-level atoms and a silver nanowire, their scattering spectra manifest double Rabi splittings associated with two atomic transitions. It was also demonstrated that plasmon-polaritons induce strong interactions between these transitions leading to Fano lineshapes of scattering resonances.

\begin{acknowledgments}
This work is partially supported by the National Science Foundation under Grant PHY-1205454 and grant from the Institute of Atomic, Molecular, and Optical Physics and the Smithsonian Center for Astrophysics.
\end{acknowledgments}

\newpage

\newpage

\newpage

\newpage

\newpage

\newpage

\end{document}